\pdfoutput=1
\documentclass[aps,prl,twocolumn,superscriptaddress,a4paper,
    floatfix]{revtex4}
\usepackage[latin1]{inputenc}
\usepackage{graphicx}
\usepackage{amsmath,amssymb}
\usepackage{hyperref}
\usepackage{datetime}
\usepackage{siunitx}
\usepackage{braket}
\usepackage{cleveref}
\usepackage{gensymb}

\begin{document}

\title{Strongly correlated Fermions strongly coupled to light}

\author{Kevin Roux}
\affiliation{Institute of Physics, EPFL, 1015 Lausanne, Switzerland}
\author{Hideki Konishi}
\affiliation{Institute of Physics, EPFL, 1015 Lausanne, Switzerland}
\author{Victor Helson}
\affiliation{Institute of Physics, EPFL, 1015 Lausanne, Switzerland}
\author{Jean-Philippe Brantut}
\affiliation{Institute of Physics, EPFL, 1015 Lausanne, Switzerland}
\date{\pdfdate}

\date{\pdfdate}

\begin{abstract}
Strong quantum correlations in matter are responsible for some of the most extraordinary properties of material, from magnetism to high-temperature superconductivity \cite{Tokura:2017aa,Basov2017aa}, but their integration in quantum devices requires a strong, coherent coupling with photons, which still represents a formidable technical challenge in solid state systems. In cavity quantum electrodynamics, quantum gases such as Bose-Einstein condensates \cite{Slama:2007ab,Colombe:2007aa,Brennecke:2007aa} or lattice gases \cite{Klinder:2015ab,Landig:2016aa} have been strongly coupled with light. However, neither Fermionic quantum matter, comparable to electrons in solids, nor atomic systems with controlled interactions, have thus far been strongly coupled with photons. Here we report on the strong coupling of a quantum-degenerate unitary Fermi gas with light in a high finesse cavity. We map out the spectrum of the coupled system and observe well resolved dressed states, resulting from the strong coupling of cavity photons with each spin component of the gas. We investigate spin-balanced and spin-polarized gases and find quantitative agreement with ab-initio calculation describing light-matter interaction. Our system offers complete and simultaneous control of atom-atom and atom-photon interactions in the quantum degenerate regime, opening a wide range of perspectives for quantum simulation.
\end{abstract}

\maketitle

Strong and coherent light-matter interactions are at the core of emerging quantum technologies, enabling the observation and control of matter at the level of single quanta \cite{Haroche:2013aa}. In many-body systems, it is reached quantitatively when the collective cooperativity $C_N=4Ng^2/\kappa \Gamma$ exceeds unity, i.e. when the fraction of photons coherently scattered into one particular mode of the electromagnetic field, singled-out by a high-finesse resonator, dominates over incoherent loss processes \cite{Tanji-Suzuki:2011ac}. Here $g$ is the coupling strength between a single photon and a single matter excitation, $N$ is the number of identical emitters, and $\kappa$ and $\Gamma$ are the incoherent decay rates of photons and matter excitations, respectively. Strong-coupling to light would be highly beneficial for the quantum simulation of interacting Fermions where first-principles theoretical calculations are inherently difficult \cite{Cirac:2012aa}. Indeed, recent theoretical work in both cold atoms and solid state systems suggests that strong coupling would make the realization and control of new quantum states of matter \cite{Kollath:2016aa,Mivehvar:2017aa,Colella:2018aa,Sheikhan:2019aa,Schlawin:2019aa,Schlawin:2019ab,Curtis:2019aa,Mazza:2019aa,Colella:2019aa} possible, as well as high-precision, quantum-limited measurements \cite{Uchino:2018ab}.

\begin{figure*}[ht]
  \includegraphics[width=1\textwidth]{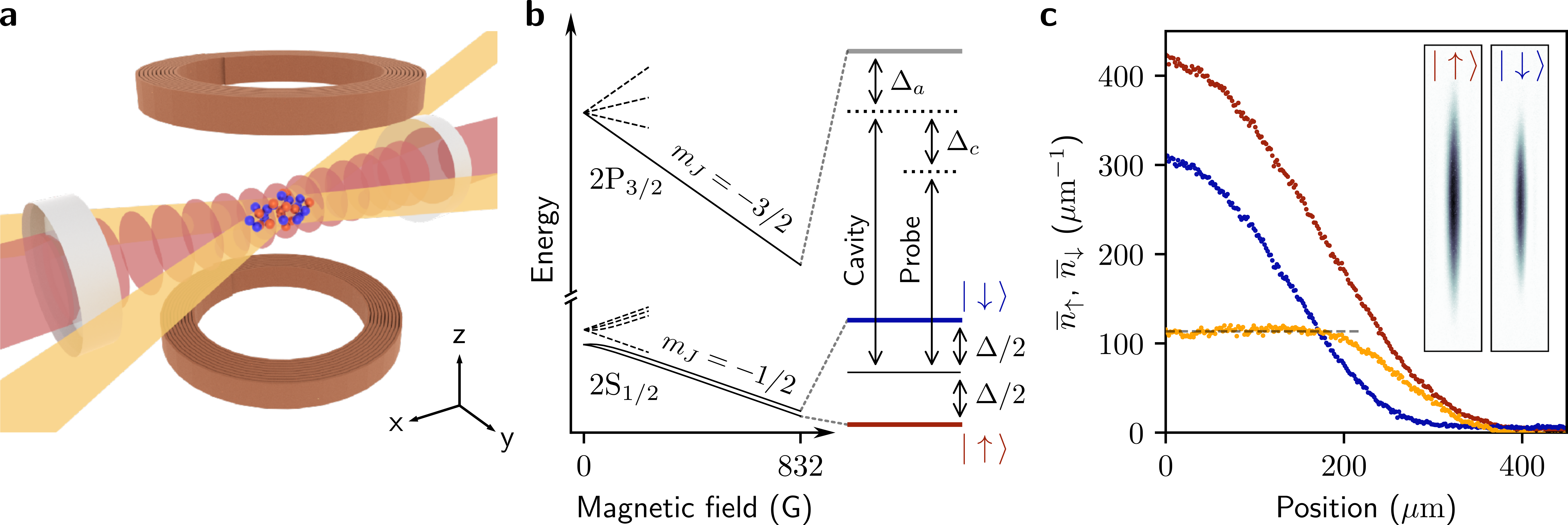}
  \caption{\textbf{Combining a unitary superfluid with a high finesse cavity.} \textbf{a,} A two-component Fermi gas is trapped in a crossed-dipole trap aligned onto the mode of a high-finesse optical cavity. A bias magnetic field oriented along the $z-$axis controls the interactions between the atoms.
  \textbf{b,} Relevant energy levels of $^6$Li at $832\,$G including the frequencies of the probe laser and cavity resonance. In the experiment the detunings $\Delta_a$ and $\Delta_c$ are varied independently. \textbf{c,} Doubly-integrated density profiles, $\overline{n}_{\uparrow}$ and $\overline{n}_{\downarrow}$, along the longitudinal direction after transfer in an elongated trap, for $9.3(5) \times 10^{4}$ and $6.0(3) \times 10^{4}$ atoms in state $\ket{\uparrow}$ (red) and $\ket{\downarrow}$ (blue) respectively. The difference between the two doubly-integrated profiles (orange) exhibits a plateau, characteristic of a superfluid core surrounded by a polarized shell. The grey dashed line is horizontal to guide the eye. Inset: absorption images ($173 \times 919 \, \si{\micro\meter}$) taken along the $z-$axis of both spin states in the elongated trap after $300 \, \si{\micro\second}$ time-of-flight. The peak column density is $430 \,\si{\micro\meter}^{-1}$.}
  \label{Fig1}
\end{figure*}

The strong coupling regime has been achieved with optical photons in various systems with weakly interacting emitters, from semiconductors and 2D material microcavities\cite{Khitrova:2006aa,Basov:2016aa}, to atoms and trapped ions \cite{Miller:2005aa,Ritsch:2013aa,Reiserer:2015aa}, including recently thermal Fermionic atoms \cite{Braverman:2019aa}. Combining evaporative cooling with high finesse cavities \cite{Slama:2007ab,Colombe:2007aa,Brennecke:2007aa,Gupta:2007aa} enabled the production of weakly interacting Bose-Einstein condensates strongly coupled with photons. Recently, Bosonic Mott insulators have been dispersively coupled to light \cite{Klinder:2015ab,Landig:2016aa}, representing the only example combining strongly correlated quantum matter and strong light-matter interactions to date.

We produce a paradigmatic example of strongly correlated system, the unitary Fermi gas, inside a high finesse cavity in the strong coupling regime. The core of the experiment is a $4.13(3)$ \si{\centi\meter}-long Fabry-Perot cavity with a finesse of $F=4.7(1)\times 10^{4}$, resonant at $671 \, \si{\nano\meter}$ with the dipole-allowed transition frequency of $^6$Li, depicted in figure \ref{Fig1}a. The experimental sequence starts with a magneto-optical trap, producing laser-cooled $^6$Li atoms within the cavity, followed by all-optical evaporation, first in a cavity-enhanced standing-wave dipole trap \cite{Mosk:2001aa}, then in a crossed dipole trap. This yields a two-component Fermi gas in the two lowest hyperfine states denoted $\ket{\uparrow}$ and $\ket{\downarrow}$ (figure \ref{Fig1}b), with tunable populations $N_{\uparrow}$ and $N_{\downarrow}$. The whole evaporation takes place under an external magnetic field of $832\,$G, the location of a broad Feshbach resonance, where atomic collisions are resonant. We obtain a degenerate unitary Fermi gas of typically $2 \times 10^{5}$ atoms in a controlled mixture of the two hyperfine states, held in a trap with an aspect ratio of three, elongated along the cavity direction (see Methods).

One of the hallmarks of strongly-interacting Fermi gases is the onset of superfluidity in the deeply degenerate regime \cite{Zwierlein:2005ab,Zwierlein:2006fk,Ketterle:2008aa}. We characterise superfluidity by observing phase separation, occurring below the critical temperature in a moderately spin-imbalanced gas \cite{Shin:2008aa}. We transfer the atoms into a single-beam trap from the crossed dipole trap by adiabatically ramping down the power of one arm (see Methods), and measure the doubly-integrated density profile difference between the two spin states along the longitudinal direction of the trap \cite{Partridge:2006fk,Shin:2006aa,Nascimbene:2010ys}, as shown in figure \ref{Fig1}c. The difference is constant at the center of the cloud, demonstrating phase separation between a superfluid, fully paired core, and a spin-polarized shell.

\begin{figure*}
  \includegraphics[width=1\textwidth]{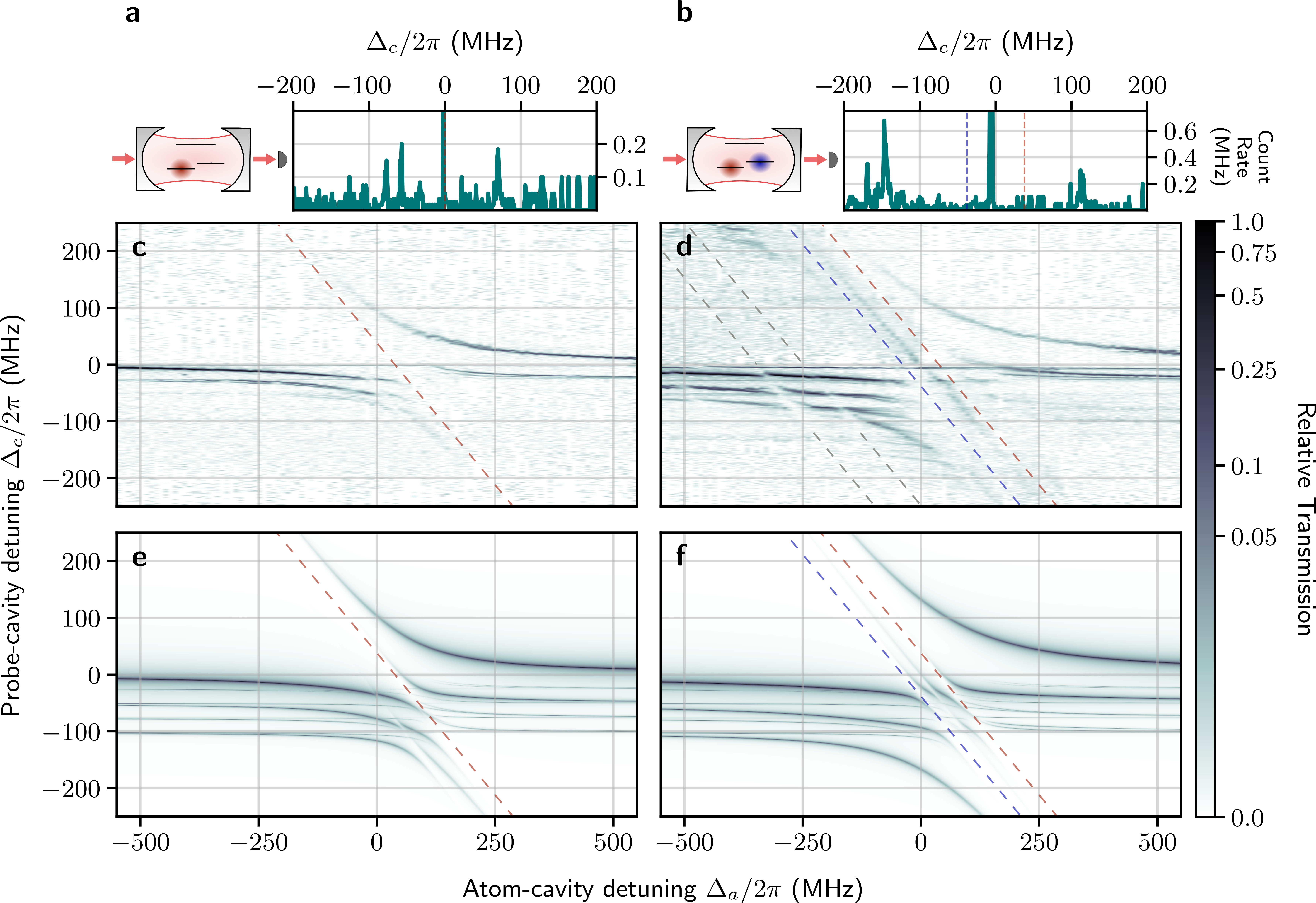}
  \caption{\textbf{Transmission spectrum of a unitary Fermi gas strongly coupled to light.}  
\textbf{a, b,} Schematic view of the experiment, the three relevant energy levels and cavity transmission signal, averaged over three realizations, as a function of $\Delta_{c}/2\pi$, for $\Delta_{a}/ 2\pi=38.2 \, \si{\mega\hertz}$ for a spin-polarized Fermi gas with $N_{\uparrow}=9.1(7)\times 10^{4}$ and $N_{\downarrow}=2(1)\times 10^{3}$ (\textbf{a}) and for $\Delta_{a}=0$ for a spin-balanced Fermi gas of $2.0(1)\times 10^{5}$ atoms equally populating each spin state (\textbf{b}). \textbf{c, d,} Coupled-system transmission spectrum as a function of $\Delta_{c}/ 2\pi$ and $\Delta_{a}/ 2\pi$ for the spin-polarized (\textbf{c}) and spin-balanced Fermi gas (\textbf{d}). The transmissions normalized by the maximum observed in each case are displayed in logarithmic scale. The bare atomic transitions for $\ket{\uparrow}$ and $\ket{\downarrow}$ are shown with red and blue dashed lines, respectively. The grey dashed lines indicate a weak contribution attributed to molecular states. \textbf{e, f,} Transmission spectra obtained from ab-initio calculation accounting for the higher-order transverse mode contributions for the spin-polarized (\textbf{e}) and the spin-balanced gases (\textbf{f}).}
  \label{Fig2}
\end{figure*}

We now investigate the coupling of the unitary Fermi gas to the cavity field, characterised by the parameters $(g,\kappa,\Gamma)=2\pi\times (0.479,0.077,5.872)$ \si{\mega\hertz} with $C_{1}=2.02$, allowing to reach the strong coupling regime for a few atoms. At 832 G, in the Paschen-Back regime, the $\sigma^-$ transition $\ket{2S_{1/2},m_{J}=-1/2} \longrightarrow \ket{2P_{3/2},m_{J}=-3/2}$ is an ideal two-level system for each spin component. The two spin states are separated by the hyperfine splitting $\Delta/ 2\pi = 76.3 \, \si{\mega\hertz}$, as depicted in figure \ref{Fig1}b together with the other relevant energy levels. The magnetic field is oriented perpendicular to the cavity axis and we probe this transition with light linearly polarized along the $y-$axis, which reduces $g$ by a factor $\sqrt{2}$, down to $g=2\pi \times 0.339$ \si{\mega\hertz}.

The general spectrum of the coupled light-matter system consists of dressed states, one for each component of the gas and each mode of the cavity. They are coherent superpositions of photonic and atomic states, with a relative weight controlled by the detuning between the atomic and the cavity resonances as well as the light-matter coupling strength. To characterize this spectrum, we inject in the cavity a weak probe beam matched with the TEM$_{00}$ mode and measure its transmission with a single photon counter, with a total detection efficiency of $0.44(5)$. We perform transmission spectroscopy as a function of the detunings $\Delta_{c}$ and $\Delta_{a}$, defined in figure \ref{Fig1}b. The probe frequency is swept at a rate of 100 \si{\mega\hertz/\milli\second} and its power is set to keep the intracavity photon number below  $\Gamma^2/8g^2=38$, the saturation threshold on resonance (see Methods).

We first measure the cavity transmission in the case of a spin-polarized Fermi gas with $N_\uparrow=9.1(7) \times 10^{4}$ and $N_\downarrow = 2(1) \times 10^{3}$. We expect the spectrum to be dominated by the contribution of the state $\ket{\uparrow}$, demonstrating the simplest case of a single component, non-interacting Fermi gas coupled to light. A typical transmission spectrum for $\Delta_{a}=0$ \si{\mega\hertz} is shown in figure \ref{Fig2}a, as a function of $\Delta_{c}/2\pi$. Figure \ref{Fig2}c shows $111$ such spectra, in logarithmic scale, for $\Delta_{a}/2\pi$ spanning $1.1 \, \si{\giga\hertz}$. We observe a prominent anti-crossing, characteristic of strong light-matter coupling, as the cavity resonance approaches the atomic one for $\ket{\uparrow}$ located at $\Delta_{a}/2\pi=38.2$ \si{\mega\hertz}. While the frequencies of the transmission peaks reveal the excitation spectrum, the value of the transmission reflects the weight of the photonic fraction in the respective excitations. For $\Delta_{c}>0$, the dressed state continuously changes from a mostly photonic excitation for large $\Delta_{a}$, to a mostly atomic excitation as $\Delta_{a}$ is decreased to negative values.

The spectrum displays a more rich structure for $\Delta_{c}<0$, due to the presence of several families of high-order cavity modes, spaced by multiples of $25$ \si{\mega\hertz} on the red side of the bare TEM$_{00}$ mode. Each of these mode families couples to the atomic resonance, producing several transmission peaks appearing in figure \ref{Fig2}a, associated to dressed states mixing high-order and TEM$_{00}$ cavity modes with the atomic excitations. Even though the coupling efficiency of the probe beam with these high-order modes is less than $10^{-3}$, the interaction with the finite size atomic cloud allows for scattering of the incident power into these modes, yielding large transmission for dressed states involving them \cite{Wickenbrock:2013aa,Kollar:2017aa}. The weaker anti-crossing centered at $\Delta_a /2\pi = -38.2\,$\si{\mega\hertz} originates from the coupling of light to the minority population in state $\ket{\downarrow}$. 

Additionally, the probe light also contains a weak linear polarization component along the magnetic field direction, which does not couple to the $\sigma^-$ transition and yields a narrow transmission peak between $\Delta_c /2\pi = 0$ and $-10 \, \si{\mega\hertz}$, as it can be seen in figure \ref{Fig2}a. This polarization component couples to the $\pi$ transition $\ket{2S_{1/2},m_{J}=-1/2} \longrightarrow \ket{2P_{3/2},m_{J}=-1/2}$, located at $\Delta_a /2\pi = 1.56\,$\si{\giga\hertz}. 

Similarly, we perform transmission spectroscopy with a spin-balanced gas of $2.0(1) \times 10^{5}$ atoms equally populating both spin states. The results are presented in figure \ref{Fig2}d, showing two anti-crossings as the bare cavity resonance approaches each atomic one, leading to three dressed state branches located above, below and between the two atomic resonances. The low relative transmission of the middle branch results from the large value of the collective Rabi frequency of approximately $2\pi \times 150$ \si{\mega\hertz} compared with the hyperfine splitting. Consequently this dressed state is mainly of atomic nature at all detunings, crossing over from state $\ket{\downarrow}$ to $\ket{\uparrow}$ as $\Delta_{a}$ goes from negative to positive.

In contrast with the non-interacting spin-polarized gas spectrum, the spin-balanced case shows not only strong coupling to the atomic resonances, but also weaker coupling to a set of matter-like excitations on the red side of the $\ket{2S_{1/2},m_{J}=-1/2} \longrightarrow \ket{2P_{3/2},m_{J}=-3/2}$ transition, as indicated in figure \ref{Fig2}d. Because of the absence of atomic transitions in this region, we attribute these excitations to molecular transitions to weakly bound states in the $2S_{1/2} + 2P_{3/2}$ asymptotic potential. A detailed study of these molecular effects is beyond the scope of this paper, but on general grounds such transitions probe the short range two-body correlations \cite{Partridge:2005aa}, which on the one hand are suppressed by the Pauli principle in the spin-polarized gas, and on the other hand are enhanced due to pairing and resonant scattering in the balanced case \cite{Junker:2008aa,Werner:2009aa}. These molecular effects have never been observed with Bose-Einstein condensates in cavities, which suggests they originate from the presence of strong-interactions.

We compare our observations with an ab-initio calculation accounting for the multimode and multilevel structure of the system. We model atoms in each state as independent two-level systems, disregarding their motional degrees of freedom, distributed in space according to the zero-temperature equation of state, coupled to several families of transverse modes (see Methods). We solve the master equation for the steady state intra-cavity field including a coherent driving of the TEM$_{00}$ mode, the atomic and cavity decays, without any free parameter. The results are shown in figure \ref{Fig2}e and \ref{Fig2}f for spin-polarized and spin-balanced gases, respectively. The model well reproduces both the location and shape of the anti-crossing of both majority and minority spin states with the TEM$_{00}$. The structure of the dressed states also agree qualitatively for the high-order modes, but the precise location and strength of the various lines strongly depend on the exact position of the cloud, which is only known with limited accuracy. The evolution of the relative transmission along the dressed state branches also agrees qualitatively with the experimental spectrum.

\begin{figure}
  \includegraphics{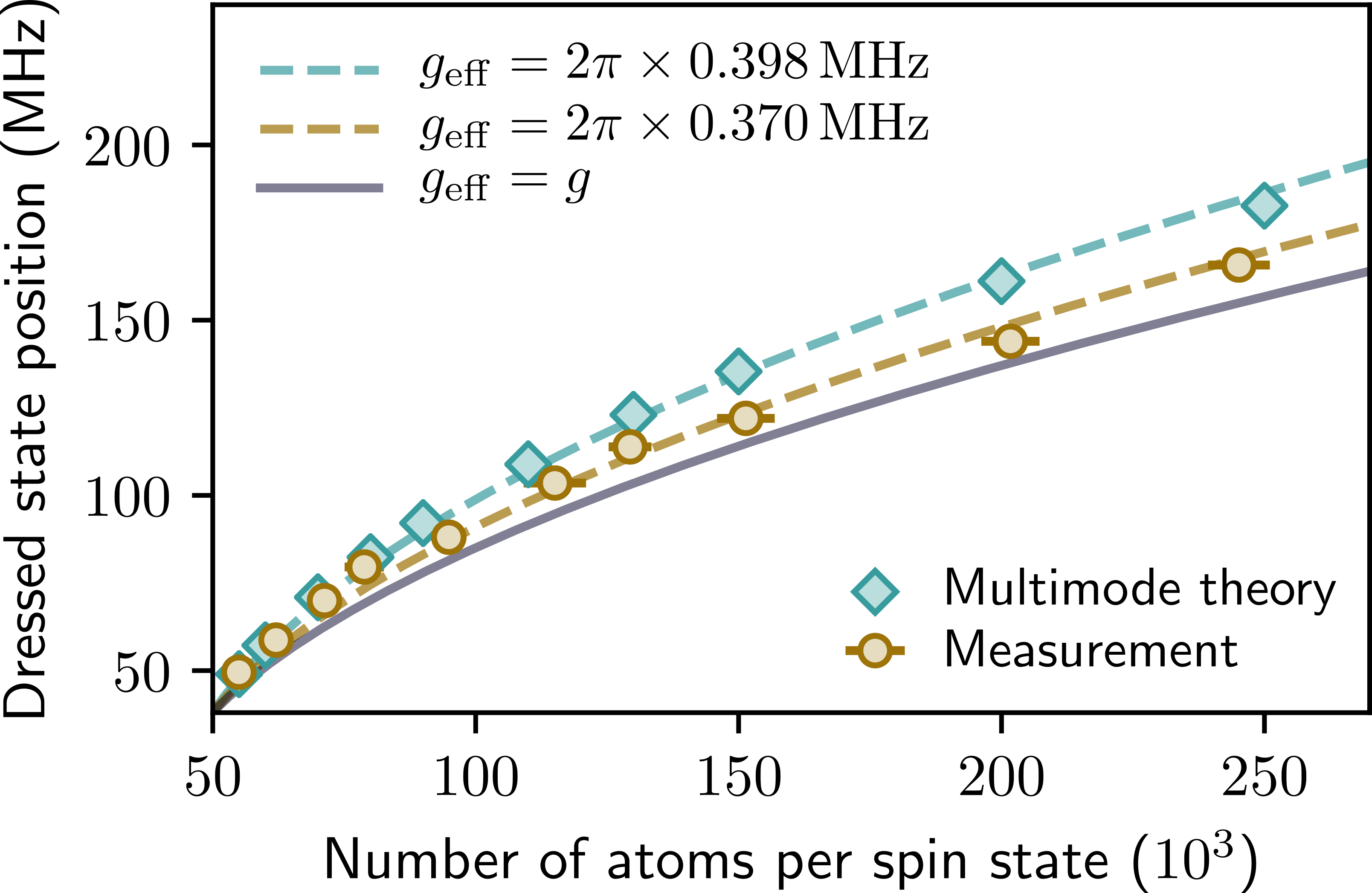}
  \caption{\textbf{Scaling of the dressed state frequency with atom number.} Position of the upper dressed state as a function of atom number in each spin state for a spin-balanced gas, with $\Delta_{c}>0$ and $\Delta_{a}=0$. Orange circles show experimental data and light blue diamonds are obtained from the ab-initio theory calculation accounting for all the high-order cavity modes. The lines describe an analytical model with two atomic states and the TEM$_{00}$ cavity mode only. The dashed lines are fits of both experimental data (orange) and theory calculation (light blue) with $g_{\mathrm{eff}}$ as a free parameter (see text). The purple solid line is the expected scaling for $g_{\mathrm{eff}}=g=2\pi \times 0.339$ \si{\mega\hertz }. Error bars are given by the statistical fluctuations over 10 realizations.}
  \label{Fig3}
\end{figure}

We now verify the coherent nature of the light-matter coupling by tracking the position of the upper dressed state for $\Delta_{a}=0$ as a function of atom number per spin state $N$ in the spin-balanced case. Atom numbers are measured independently by absorption imaging along the $z-$axis. The results and the predictions of the multimode model are presented together in figure \ref{Fig3}. The difference between the model and the data is at most $8\%$. We fit both the experimental data and the theoretical model by $\sqrt{g_{\mathrm{eff}}^{2}N+\frac{\Delta^{2}}{4}}$, with $\Delta$ the hyperfine splitting (lines in figure \ref{Fig3}), describing two equally populated atomic states coherently and uniformly coupled to a single cavity mode (see Methods). 
Leaving $g_{\mathrm{eff}}$ as an adjustable parameter, both the data and theory are well fit, confirming the coherent nature of light-matter coupling. This yields $g_{\mathrm{eff}}=2\pi \times 0.370$ and  $2\pi \times 0.398$ \si{\mega\hertz}, $8\%$ and $16\%$ larger than $g$ respectively. We attribute these differences to the role of high-order modes, which is slightly overestimated in the ab-initio model (see Methods).

A Fermi gas with both strong light-matter and atom-atom interactions opens many unexplored directions. In the dispersive regime, the cavity mediates a long-range interaction \cite{Ritsch:2013aa}, making this system an ideal platform to explore the interplay between Fermi statistics, density wave ordering \cite{Piazza:2014aa,Chen:2014ab,Keeling:2014aa} and superfluidity \cite{Chen:2015aa}, a paradigm of competing orders also encountered in strongly correlated materials \cite{Sachdev:2000aa}. The cavity provides a controlled dissipation channel, a promising tool to prepare new correlated states of Fermions \cite{Muller:2012ab}. Correspondingly, the photon leakage realizes a weak continuous measurement whose back-action can compete with short-range interactions \cite{Mazzucchi:2016ab} and allow for the implementation of feedback protocols \cite{Mazzucchi:2016aa}. Lastly, our evidence for molecular transitions addressed by cavity photons offers the possibility to control interactions \cite{Theis:2004aa,Bauer:2009aa} using dynamical, quantized fields, opening a new class of models to quantum simulation.

\section*{Acknowledgements}

We thank Tilman Esslinger, Tobias Donner, Tigrane Cantat-Moltrecht and Christophe Galland for discussions and reading of the manuscript, Barbara Cilenti for help in the early stages of the experiment, Ilaria Di Meglio and Nikolai Klena for proof-reading the manuscript and Martin Lebrat for help with the spectroscopic data. We acknowledge funding from the European Research Council (ERC) under the European Union's Horizon 2020 research and innovation programme (grant agreement No 714309), the Swiss National Science Foundation (grant No 184654), the Sandoz Family Foundation-Monique de Meuron program for Academic Promotion and EPFL.

\section*{Author contributions}

All authors contributed extensively to the work presented here.

\section*{Methods}

\subsection*{Preparation of a degenerate unitary Fermi gas in a high finesse cavity}

The experimental sequence starts with approximately $4 \times 10^8$ $^6$Li atoms captured in a magneto-optical trap (MOT). After $3$ \si{\second} loading, the MOT is compressed, optically pumped into the $F=1/2$ hyperfine manifold, and about $10^7$ laser-cooled atoms are loaded into a cavity-enhanced standing-wave dipole trap at $1064$ \si{\nano\meter}. The cavity has a finesse of $\sim 3000$ at 1064 \si{\nano\meter}, providing a power build-up of $\sim1000$. We use the TEM$_{10}$ cavity mode with a vertical nodal line for dipole trapping, in order to mitigate the thermal effects of the $1064$ \si{\nano\meter} light on the cavity stability.

The bias magnetic field is ramped up to $832$ G within $250$ \si{\milli\second}, before the evaporative cooling starts. The first evaporation ramp in the standing-wave trap lasts $300$ \si{\milli\second}. Approximately $1.5 \times 10^6$ atoms are then transferred in the running-wave crossed dipole trap by turning off the standing-wave dipole trap. The crossed dipole trap is made of two non-interfering, $1.5$ \si{\watt}, $1064$ \si{\nano\meter} laser beams focused to waists of $32(1)$ \si{\micro\meter} and intersecting at the cavity mode waist, with an angle of $35\degree$. A $350$ \si{\milli\second} long linear ramp reduces the power down to $80 \, \si{\milli\watt}$ which completes the evaporation, then followed by a recompression stopping the evaporation and maximizing the coupling with the $45.0(3)$ \si{\micro\meter} waist TEM$_{00}$ cavity mode at $671$ \si{\nano\meter}. For the measurements, the trapping frequencies are $(\omega_x,\omega_y,\omega_z)/2\pi=(300(30),924(40),902(11))$ \si{\hertz}, yielding a Fermi energy of $E_{\mathrm{F}}/h=53(5)$ \si{\kilo\hertz} for $N_\uparrow = N_\downarrow=1.00(5) \times 10^5$, where $h$ is the Planck constant.

The populations in states $\ket{\uparrow}$ and $\ket{\downarrow}$ are tuned during the evaporation process. Balancing the populations between the two spin states is achieved using incomplete Landau-Zener radio-frequency (RF) sweeps during the entire evaporation, and the total atom number can be varied from $1 \times 10^3$ to $4 \times 10^5$ by adjusting the evaporation endpoint. A controlled spin-imbalance is introduced by removing the RF sweeps and introducing losses in the $\ket{\downarrow}$ population using the $p$-wave Feshbach resonance located at $214.9$ G before evaporation. The largest population difference corresponds to $N_{\uparrow}=9.1(7) \times 10^4$ and $N_{\downarrow}=2(1) \times 10^3$.

The length of the cavity is stabilized during the entire experimental sequence using the Pound-Drever-Hall stabilization technique with an additional laser at $532$ \si{\nano\meter}, for which the cavity has a finesse of approximatively $3000$, on the TEM$_{02}$ mode. The power injected in the cavity creates a peak lattice potential of $46$ \si{\nano\kelvin}$=8\times10^{-3} E_R/k_B$, with $E_R$ the recoil energy and $k_B$ the Boltzmann constant.

\subsection*{Superfluidity characterization}

In our crossed dipole trap, the large density and small cloud size preclude in-situ imaging, thus preventing a direct measurement of the equation of state. To circumvent this problem, we transfer the Fermi gas from the crossed dipole trap to a single beam dipole trap by adiabatically turning off one of the arms after the entire evaporation is completed. The atoms are then held in the elongated trap for another $250$ \si{\milli\second} to ensure thermal equilibrium. This trap has trapping frequencies $2\pi \times (28.5(2),846(4),846(4))$ \si{\hertz}, with longitudinal trapping ensured by the magnetic field curvature. We image the atoms in either spin state after a time-of-flight of $300$ \si{\micro\second}, short compared with the longitudinal trap frequency, which increases the transverse size above the resolution of our imaging system.
Integration along the transverse direction yields figure \ref{Fig1}c.

\subsection*{Multimode theoretical model}

The model contains two sets of two-level systems describing the two hyperfine states $\ket{\uparrow}$ and $\ket{\downarrow}$, with resonance frequencies $(\omega_{e,\uparrow},\omega_{e,\downarrow})/2\pi$. We model the high order modes of the cavity by four families of TEM$_{mn}$ modes, with $n+m\in\{11,22,33,44\}$, in addition to the TEM$_{00}$, so that $115$ modes are accounted for in total. Successive families are spaced by $25$ \si{\mega\hertz}, and all modes within one family are degenerate. We designate the mode frequencies by $\omega_\nu/2\pi$, $\nu$ labeling each individual mode. 

Introducing the annihilation operators $\hat{a}_\nu$ for the cavity modes and the Pauli matrices $\hat{\sigma}_z^{(i,\lambda)}$ for atom $i$ in state  $\lambda=\uparrow, \, \downarrow$, the free Hamiltonian reads (with $\hbar=1)$:
\begin{equation}
\hat{H}_0 = \sum_\nu \omega_\nu \hat{a}_\nu^\dagger \hat{a}_\nu +  \sum_{i,\lambda} \omega_{e,\lambda}\frac{1 + \hat{\sigma}_z^{(i,\lambda)}}{2}.
\end{equation}
We describe the light-matter coupling within the rotating wave approximation as
\begin{equation}
\hat{H}_\mathrm{int} = -i \frac{\Omega_0}{2}\sum_{\nu,i,\lambda} \left( \hat{a}_\nu f_\nu(r_i)\hat{\sigma}_+^{(i,\lambda)}  - hc\right)
\end{equation}
where $hc$ stands for Hermite conjugate, and we have introduced the Rabi frequency $\Omega_0$ for the TEM$_{00}$ mode, calculated from spectroscopic data of $^6$Li and the Breit-Rabi formula. $r_i$ is the position of atom $i$ considered here as a fixed, classical parameter, and $f_\nu(r)$ are the mode functions, defined as $\sqrt{V} u_\nu(r)$, $V$ is the mode volume common to all modes, and the functions $u_\nu$ are an orthonormalized set. We also include the external driving at frequency $\omega/2\pi$:
\begin{equation}
\hat{H}_\mathrm{drive} = i\sqrt{\kappa}\sum_\nu \mathcal{F}_\nu \left( \hat{a}_\nu e^{-i\omega t} - \hat{a}_\nu^\dagger e^{i\omega t} \right)
\end{equation}
where the driving strength of mode $\nu$ is $\mathcal{F}_\nu$.

We then search for the steady state solution of the master equation including cavity decay and spontaneous emission. We neglect quantum fluctuations and atom-field correlations, thus replacing the field operators by coherent amplitudes $\alpha_\nu$ and $\left\langle\hat{\sigma}_{z,\pm}^{(i,\lambda)}\right\rangle$. In the low saturation approximation, we further have $\left\langle\hat{\sigma}_{z}^{(i,\lambda)}\right\rangle \sim -1$ and $\left\langle\hat{\sigma}_{-}^{(i,\lambda)}\right\rangle \sim \frac{\Omega_0/2}{\Gamma/2 +i\Delta_\lambda}$, with $\Delta_\lambda$ the detuning between the driving and the resonance for state $\lambda$.

This way, we eliminate the atomic degrees of freedom and obtain a set of algebraic equations for the field amplitude in mode $\nu$: \begin{multline}
- \frac{\Omega_0^2}{4} \sum_{\mu} \left(\sum_\lambda  \frac{\int_V d^3r n_\lambda(r) f_\nu(r)f_\mu(r)}{\frac{\Gamma}{2}+i\Delta_\lambda} \right) \alpha_\mu - \sqrt{\kappa}\mathcal{F}_\nu \\= (i \delta_\nu + \frac{\kappa}{2}) \alpha_\nu
\end{multline}
where the index $\mu$ runs over all the modes, and we have replaced summation over the positions of the atoms by an integral over the density distribution $n_\lambda(r)$ of atoms in state $\lambda$. This highlights the essential role of the finite size of the cloud in redistributing the photons between different cavity modes. 

The total light intensity in the cavity $\sum_\nu |\alpha_\nu|^2$ is shown in figure \ref{Fig2}. In the linear regime where the model is valid this is equivalent to the outgoing photon flux up to a trivial normalization. 

The overlap integrals with each of $115$ modes are calculated using the equilibrium, zero temperature equation of state of the unitary and ideal Fermi gases for the balanced and highly polarized cases, respectively. We suppose that the cloud is perfectly centered on the cavity axis, and we use ideal Hermite-Gauss modes. 

While the coupling with the TEM$_\mathrm{00}$ mode is weakly sensitive to small misalignments and mode shape imperfections, the predictions for higher order modes are much more sensitive. In particular, the detailed redistribution of photons among the modes changes upon varying the position of the cloud by a fraction of the cloud size. We also observed deviations of the high order mode profiles compared with ideal Hermite-Gauss modes, such as distortion of the nodal lines, which could be due to cavity mirrors misalignments. As a result, the mode volume for higher transverse modes is likely larger than theory would predict, a possible source of the overestimation of their role in the spectrum.

\subsection*{Analytical model}

To provide further physical intuitions, we also compared our data with an analytical model, generalizing the Tavis-Cummings model to a balanced mixture of two independent internal states with resonance frequencies $(\omega_{e,\uparrow},\omega_{e,\downarrow})/2\pi$, with total atom number $2N_0$. The model's Hamiltonian reads
\begin{multline}
\hat{H} = \omega_{e,\uparrow} \hat{J}_{z,\uparrow} + \omega_{e,\downarrow} \hat{J}_{z,\downarrow} + \omega_0 \hat{a}^\dagger \hat{a} \\+ g_{\mathrm{eff}} \left(  \hat{J}_{+,\uparrow} \hat{a} + \hat{J}_{+,\downarrow} \hat{a} + hc\right)
\end{multline}
with $\omega_0$ the resonance frequency of the cavity mode and $g_{\mathrm{eff}}$ the light-matter coupling strength. Here the $\hat{J}_{\mu,\lambda}$, $\mu=z,\pm$ are collective spin operators for atoms in state $\lambda$ \cite{Haroche:2013aa}. In the low saturation regime, the Holstein-Primakoff transformation allows to rewrite the Hamiltonian in terms of Bosonic operators $\hat{b}_{\lambda}$ describing individual, non-interacting collective optical excitations shared among atoms in state $\lambda$:
\begin{multline}
\hat{H} = \omega_0 \hat{a}^\dagger \hat{a} + \omega_{e,\uparrow} \hat{b}_\uparrow^\dagger \hat{b}_{\uparrow}+ \omega_{e,\downarrow}\hat{b}_\downarrow^\dagger\hat{b}_{\downarrow}  \\+ g_{\mathrm{eff}} \sqrt{N_0} \left(  \hat{b}_{\uparrow}^\dagger \hat{a} +  \hat{b}_{\downarrow}^\dagger \hat{a} + hc\right)
\end{multline}
where constant terms have been dropped. For $\omega_0 - \omega_\uparrow = -(\omega_0 -\omega_\downarrow) = \Delta/2$, the normal modes of this coupled harmonic oscillator model have frequencies $E_{0,\pm} = 0,\pm \sqrt{2 g_{\mathrm{eff}}^2 N_0 + \Delta^2/4}$. 

This supposes that all the atoms are maximally coupled to the field. In practice, due to the cosine longitudinal mode shape, the number of atoms coupled to the field is reduced by a factor of $2$, yielding the fit function used in the main text. 

\subsection*{Cavity transmission spectroscopy}

The probe laser is locked onto a transfer cavity and narrowed-down to a linewidth $\leq 10 \, \si{\kilo\hertz}$ using the Pound-Drever-Hall stabilization technique. Its absolute frequency is regulated using a wavemeter referenced onto a laser frequency-stabilized by saturated absorption spectroscopy on the $\ket{2S_{1/2}} \longrightarrow \ket{2P_{3/2}}$ transition of $^{6}$Li. 

For one realization of the experiment, we fix $\Delta_a $ and sweep the probe laser frequency by $\pm25 \, \si{\mega\hertz}$ within $500 \, \si{\micro\second}$, using a broadband acousto-optic modulator, thereby covering a $50 \, \si{\mega\hertz}$ range in $\Delta_c/2 \pi$. A fast sweep rate is necessary in order to minimize the effect of atomic motion during the measurement. Over such a scan the probe power varies by at most $10\%$. The detected signals are averaged over three realizations. The frequency sweep rate is larger than $\kappa^2/2\pi$, so that the field will not reach the steady state in the absence of atoms. However, based on the analytical model we expect the decay rate for the dressed state \begin{equation}
\frac{\kappa + \Gamma\Omega_{0}^{2}/2\Delta_a^2}{1+\Omega_{0}^{2}/2\Delta_a^2}
\end{equation}
to be dominated by the atomic decay rate for all the parameters covered in the experiment. This ensures that steady state conditions are realized during a measurement. This may however not be the case for weakly coupled higher order modes and for the weak $\pi$ polarization contribution.

We estimate the average intracavity photon number, for the maximal count rate detected with the cavity set resonant with either of the two atomic states. It assumes the steady state, and loss in the mirrors inferred from a comparison between the measured finesse and the independently measured transmission of the mirrors. This leads to a mirror transmission of $40$ ppm and loss of $25$ ppm yielding an average intracavity photon number of $10$. This ensures we work below saturation.

\bibliography{paper}
\bibliographystyle{naturemag}

\end{document}